\documentclass[11pt,a4paper]{article}
\usepackage[margin=1in]{geometry}
\usepackage{amsmath,amssymb}
\usepackage{booktabs,multirow,enumitem}
\usepackage{tikz,pgfplots,subcaption}
\pgfplotsset{compat=1.17}
\usepackage[hidelinks]{hyperref}
\usepackage{cite}

\newcommand{\repourl}{https://github.com/Deployment-inc/deep-perceptual-compute-allocation}
\newcommand{\releaseurl}{https://github.com/Deployment-inc/deep-perceptual-compute-allocation/tree/b0394458e38446b9b16cf60ca0153d0998827c0d}
\title{\textbf{Hearing's Blind Spots:\\Toward Perceptual Compute Allocation\\in Neural Text-to-Speech}}
\author{\textbf{Aayush Gupta}\\Deployment Inc.\\{\footnotesize\texttt{aayush@deployment.inc}}}
\date{October 2026}
\begin{document}
\maketitle
\begin{abstract}
A speech synthesizer produces a detailed waveform, but a listener need not hear every detail. Strong sounds can mask weaker neighbors, creating context-dependent ``blind spots'' in hearing. Could neural text-to-speech (TTS) spend less computation on those details? We investigate a first step: how the placement of spectral degradation affects completed speech from Piper and Kokoro. At matched time--frequency bin-removal budgets, a simple low-energy-first policy achieves higher mean wideband PESQ than random deletion and the tested psychoacoustic masking criteria. At 50\% removal in the spreading-matrix comparison, energy, masking, and random policies score 4.46, 3.60, and 1.19 on Piper, and 3.42, 2.10, and 1.22 on Kokoro. Spectrograms and masking maps expose where these perturbations fall and why a model's below-threshold regions cannot automatically be treated as inaudible. Calibration diagnostics show substantial threshold sensitivity, while a small single-listener pilot establishes no transparent removal budget. CPU profiles identify waveform decoding as the main measured kernel cost, but a tested dynamic-INT8 Kokoro decoder runs slower than FP32 on both CPUs. The results establish energy ranking as a strong oracle baseline and motivate testing perceptual allocation inside a decoder. They do not yet demonstrate saved computation: the decisions here use already synthesized audio, and quality metrics are evaluated at 16\,kHz. We release scripts, results, audio examples, and a listening-study instrument.
\end{abstract}

\section{Introduction}
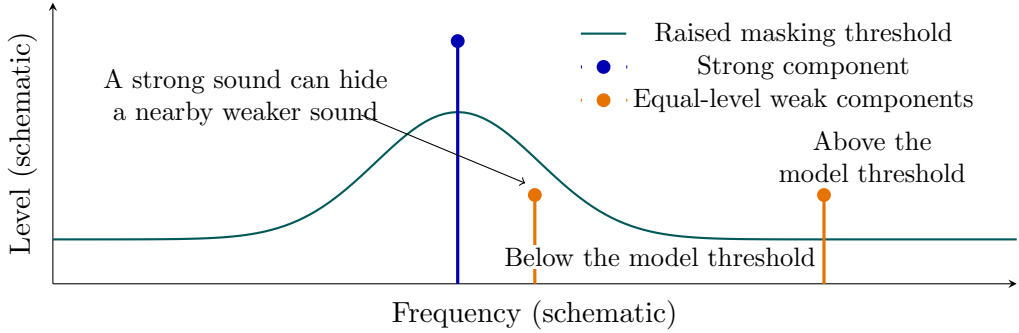
\begin{figure}[tbp]
\centering
\begin{tikzpicture}
\begin{axis}[width=.9\textwidth,height=5.3cm,xmin=0,xmax=10,ymin=0,ymax=95,
 xlabel={Frequency (schematic)},ylabel={Level (schematic)},xtick=\empty,ytick=\empty,
 axis lines=left,clip=false,legend style={font=\small,draw=none},legend pos=north east]
\addplot[color=teal!70!black,no marks,thick] coordinates {(0,15) (0.066667,15) (0.13333,15) (0.2,15.001) (0.26667,15.001) (0.33333,15.001) (0.4,15.002) (0.46667,15.003) (0.53333,15.004) (0.6,15.005) (0.66667,15.007) (0.73333,15.01) (0.8,15.014) (0.86667,15.019) (0.93333,15.026) (1,15.035) (1.0667,15.047) (1.1333,15.063) (1.2,15.083) (1.2667,15.109) (1.3333,15.143) (1.4,15.186) (1.4667,15.24) (1.5333,15.308) (1.6,15.393) (1.6667,15.499) (1.7333,15.629) (1.8,15.788) (1.8667,15.981) (1.9333,16.213) (2,16.492) (2.0667,16.823) (2.1333,17.215) (2.2,17.674) (2.2667,18.208) (2.3333,18.825) (2.4,19.532) (2.4667,20.338) (2.5333,21.248) (2.6,22.268) (2.6667,23.402) (2.7333,24.654) (2.8,26.024) (2.8667,27.511) (2.9333,29.112) (3,30.819) (3.0667,32.623) (3.1333,34.513) (3.2,36.472) (3.2667,38.483) (3.3333,40.523) (3.4,42.571) (3.4667,44.599) (3.5333,46.581) (3.6,48.488) (3.6667,50.292) (3.7333,51.965) (3.8,53.478) (3.8667,54.807) (3.9333,55.928) (4,56.822) (4.0667,57.472) (4.1333,57.867) (4.2,58) (4.2667,57.867) (4.3333,57.472) (4.4,56.822) (4.4667,55.928) (4.5333,54.807) (4.6,53.478) (4.6667,51.965) (4.7333,50.292) (4.8,48.488) (4.8667,46.581) (4.9333,44.599) (5,42.571) (5.0667,40.523) (5.1333,38.483) (5.2,36.472) (5.2667,34.513) (5.3333,32.623) (5.4,30.819) (5.4667,29.112) (5.5333,27.511) (5.6,26.024) (5.6667,24.654) (5.7333,23.402) (5.8,22.268) (5.8667,21.248) (5.9333,20.338) (6,19.532) (6.0667,18.825) (6.1333,18.208) (6.2,17.674) (6.2667,17.215) (6.3333,16.823) (6.4,16.492) (6.4667,16.213) (6.5333,15.981) (6.6,15.788) (6.6667,15.629) (6.7333,15.499) (6.8,15.393) (6.8667,15.308) (6.9333,15.24) (7,15.186) (7.0667,15.143) (7.1333,15.109) (7.2,15.083) (7.2667,15.063) (7.3333,15.047) (7.4,15.035) (7.4667,15.026) (7.5333,15.019) (7.6,15.014) (7.6667,15.01) (7.7333,15.007) (7.8,15.005) (7.8667,15.004) (7.9333,15.003) (8,15.002) (8.0667,15.001) (8.1333,15.001) (8.2,15.001) (8.2667,15) (8.3333,15) (8.4,15) (8.4667,15) (8.5333,15) (8.6,15) (8.6667,15) (8.7333,15) (8.8,15) (8.8667,15) (8.9333,15) (9,15) (9.0667,15) (9.1333,15) (9.2,15) (9.2667,15) (9.3333,15) (9.4,15) (9.4667,15) (9.5333,15) (9.6,15) (9.6667,15) (9.7333,15) (9.8,15) (9.8667,15) (9.9333,15) (10,15)};
\addlegendentry{Raised masking threshold}
\addplot[ycomb,very thick,blue!70!black,mark=*] coordinates {(4.2,82)};
\addlegendentry{Strong component}
\addplot[ycomb,very thick,orange!90!black,mark=*] coordinates {(5,30) (8,30)};
\addlegendentry{Equal-level weak components}
\node[align=center,font=\small] at (axis cs:2,63) {A strong sound can hide\\a nearby weaker sound};
\draw[->] (axis cs:3.2,57)--(axis cs:4.9,34);
\node[align=center,font=\small,fill=white,inner sep=1pt] at (axis cs:6.3,9) {Below the model threshold};
\node[align=center,font=\small] at (axis cs:8.5,43) {Above the\\model threshold};
\end{axis}\end{tikzpicture}

\caption{The intuition behind hearing's blind spots. A strong component can raise the detection threshold for nearby weaker components. The two orange components have the same level but different masking contexts. This drawing is a schematic, not a measured threshold or a result from our listener pilot.}
\label{fig:intuition}
\end{figure}

Why spend the same synthesis effort on a detail a listener can detect and one hidden by another sound? A strong component can make a weaker neighbor harder to hear. These context-dependent gaps in detectability are what we call hearing's ``blind spots'' (Figure~\ref{fig:intuition}). The phrase describes auditory masking, not a fixed missing frequency range or a listener's hearing impairment. The same weak component can become detectable when its masking context changes.

Conventional neural waveform decoders nevertheless apply a fixed network structure across output positions. This motivates \emph{perceptual compute allocation}: reserving detailed synthesis for perceptually consequential content and approximating less consequential content more cheaply. Whether this reduces latency depends on the decoder, hardware, routing overhead, and granularity of the work that can actually be skipped.

Perceptual audio codecs already exploit auditory masking to allocate bits~\cite{iso11172,painter2000perceptual,zwicker2007psychoacoustics}. Psychoacoustic criteria have also been used in neural-audio training losses~\cite{zhen2018psychoacoustically,zhen2020calibration}. Perceptual sensitivity has also motivated giving more synthesis resources to low-frequency subbands in an LPCNet-based vocoder~\cite{cui2020subband}. These precedents motivate asking whether the changing audibility of content could guide runtime synthesis decisions. Foveated rendering provides a related example of allocating work according to sensory sensitivity~\cite{guenter2012foveated,patney2016foveated,kaplanyan2019deepfovea}.

We examine a prerequisite for this idea: how does the placement of a fixed amount of output degradation affect reference-based quality scores? Our oracle harness uses the fully synthesized waveform to choose perturbations. It therefore tests output sensitivity after synthesis, not the feasibility or cost of an online allocation policy. In particular, a fraction of removed spectral bins is neither a fraction of decoder operations nor a measured compute saving.

The strongest finding is a baseline result. Energy-ranked deletion exceeds the mean PESQ of the tested masking-based criteria at each reported removal budget, and both exceed random deletion. The size of the advantage depends on the model and budget. At 50\% removal, Piper retains a mean PESQ of 4.46 under energy ranking, whereas Kokoro falls to 3.42. These differences caution against a single removal percentage applying to all TTS systems. Human transparency remains unresolved: the author-only listening pilot is too limited to establish it.

Our contributions are:
\begin{enumerate}[leftmargin=*]
\item An oracle comparison of energy, masking, and random deletion policies in two TTS model families, with matched spectral-bin budgets and explicit limits on the inference that can be drawn from PESQ.
\item Calibration diagnostics and operator-design observations that help explain why model-declared masking fractions should not be treated as perceptually removable output.
\item CPU profiling and an implementation-specific negative result for dynamic-INT8 decoder quantization, separating output redundancy from realized computation savings.
\item A release of scripts, result files, and a listening-study instrument for reproducible follow-up work at \url{\repourl}.
\end{enumerate}

\section{Background and Research Questions}
A psychoacoustic model (PAM) estimates how signal-dependent spectral and temporal masking, together with an absolute hearing threshold, affect the detectability of other components~\cite{iso11172,zwicker2007psychoacoustics,bosi2003introduction}. We write its threshold as $T(t,f)$ and define headroom as signal level minus threshold on a compatible decibel scale. Negative headroom is a \emph{model prediction}; it does not establish inaudibility for a particular listener, playback level, or simultaneous set of deletions.

The original research questions concerned (H1) perceptually redundant output, (H2) reduced end-to-end CPU time at transparent quality, and (H3) whether masking-guided allocation improves quality over energy and random allocation at matched compute budgets. The experiments here address narrower, exploratory questions: whether placement improves metric scores at matched \emph{output-removal} budgets, how sensitive the masking estimates are to calibration, and what limits arise in one candidate reduced-precision implementation. H1 remains unresolved as a quantitative human-perception claim; H2 is not demonstrated; and the oracle comparison provides no support for the anticipated masking advantage in H3.

The repository includes planning and amendment documents. We treat them as research provenance, not evidence of an independently timestamped preregistration. The amendment records changes after early results; this paper does not characterize its revised framing or future listening-study analysis as preregistered confirmatory research. Transparency, wherever discussed as a future target, means perceptual indistinguishability from the full-compute output of the same model, text, and seed.

\section{Testbed and Method}
\subsection{Models, hardware, and corpus}
We use the CPU execution provider of ONNX Runtime~\cite{onnxruntime}. The Piper~\cite{piper} voice is \texttt{en\_US-lessac-medium}, a VITS~\cite{kim2021vits} system with a HiFi-GAN decoder~\cite{kong2020hifigan} at 22.05\,kHz. Kokoro v1.0~\cite{kokoro}, using voice \texttt{af\_sarah}, has a StyleTTS2-related architecture~\cite{li2023styletts2} with an iSTFTNet decoder~\cite{kaneko2022istftnet} at 24\,kHz. The reported hardware is an Apple M5 Pro (aarch64) and an Intel Xeon Platinum 8488C (Sapphire Rapids, AVX-VNNI, eight vCPUs on AWS \texttt{c7i.2xlarge}).

The corpus contains \textbf{45 texts}: 15 conversational-agent prompts with digits and entities, 15 narration passages, and 15 romanized Hindi--English (Hinglish) prompts. Synthesizing each text with both models produces \textbf{90 waveforms}. Hinglish is processed with English voices and English grapheme-to-phoneme models; these results do not establish performance for native Hindi synthesis. Subsets are described for each experiment. The spreading-matrix deletion experiment uses the first eight texts per domain (24 per model); the noise-shaping and mel-resolution experiments use the first five (15 per model).

\subsection{Psychoacoustic implementations}
We compare a from-scratch Model-1-lineage implementation~\cite{iso11172,bosi2003introduction}, including tonal/noise maskers, an absolute-threshold approximation~\cite{terhardt1979pitch}, and forward masking, with a spreading-matrix implementation adapted from reference audio-coding code~\cite{schuller2020filter}. The analysis uses a 1024-sample Hann window and 512-sample hop. Model-1 inclusive and leave-self-out variants are variants of one implementation, not independent validated listener models. The later spreading-matrix comparison uses a global $-16$\,dB threshold offset. Section~\ref{sec:probe} distinguishes this engineering setting from perceptual calibration.

\subsection{Oracle deletion and noise shaping}
\label{sec:harness}
\begin{figure}[tbp]
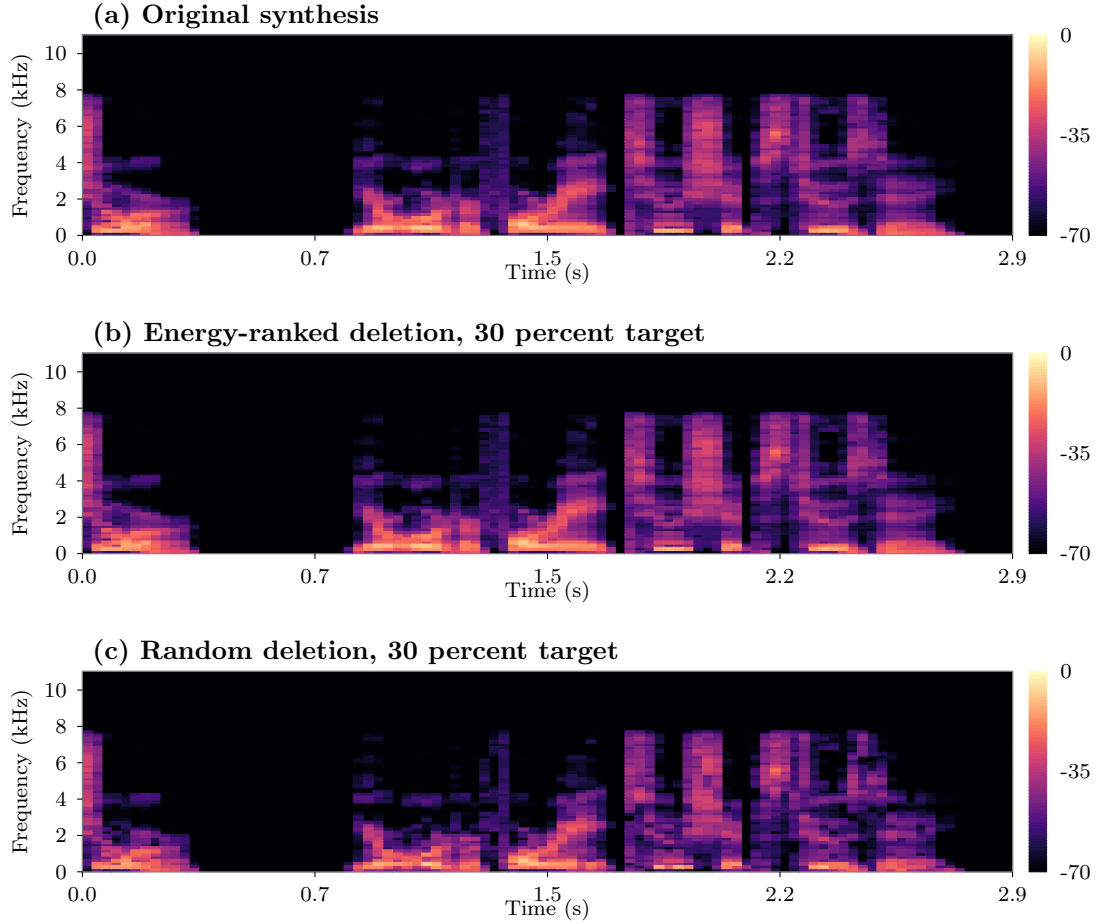

\centering

\caption{Where deletion lands matters. Native-rate spectrograms of the released Piper listening example, ``Sure, your order 4 8 2 ships Tuesday.'' All panels use the same display scale in dB relative to the reference STFT peak. Display pixels average power over neighboring bins and frames. The processed files are the released 30\% energy and random conditions, not newly synthesized examples. Visible similarity is not a test of audibility; these plots show how the perturbations differ.}
\label{fig:spectra}
\end{figure}

The deletion experiment ranks (frame, Bark-band) cells and zeros selected STFT bins before waveform reconstruction. Policies rank by increasing band energy, increasing signal-minus-exclusive-threshold headroom, or a fixed-seed random permutation. The budget is the targeted fraction of time--frequency bins removed. Bark bands have unequal widths, so cells are accumulated by their bin counts rather than by cell count. This is a discrete output-degradation budget, not an equal-compute constraint. Random deletion is a control for placement, not a baseline synthesis architecture. Figure~\ref{fig:spectra} makes this distinction visible in a released listening example: the two policies target the same nominal removal fraction but distribute it differently across the spectrum.

The separate noise experiment matches \emph{total allocated spectral noise power}, expressed as a noise-to-signal ratio (NSR). It distributes this power in proportion to signal power (energy), to the Model-1 threshold power (masking), or equally over bins (uniform). Thus noise shaping and deletion use different budget definitions. Neither perturbation reproduces all errors that a cheaper decoder or quantized kernel would produce.

Quality is scored against the original waveform with wideband PESQ~\cite{itu_p862,rix2001pesq} and STOI~\cite{taal2011stoi}, after resampling both signals to 16\,kHz. PESQ is the primary reported comparison; it does not establish transparency or intelligibility on its own. In particular, content above 8\,kHz in the native-rate waveforms is outside the scored bandwidth, while the deletion budget is defined on the native-rate spectral grid. The results therefore need a bandwidth-matched control before being interpreted as full-band redundancy. The reported unmodified analysis/synthesis round trip has 232\,dB SNR and PESQ approximately 4.64; this checks the numerical reconstruction path only.

\subsection{Criterion/operator matching and self-masking}
\label{sec:traps}
A deletion criterion concerns whether existing content is dispensable; a noise criterion concerns whether added error is detectable. Interchanging them can bias a comparison. A second issue is self-masking: a component may contribute to the threshold used to justify its own deletion. The reported deletion comparisons use leave-self-out thresholds. This mitigates self-masking but does not guarantee that all selected cells remain masked when removed together, because different cells can mask one another. Threshold estimates and metric scores remain hypotheses about perception pending listening tests.

\section{Results}
\subsection{CPU cost distribution}
\label{sec:rq2}
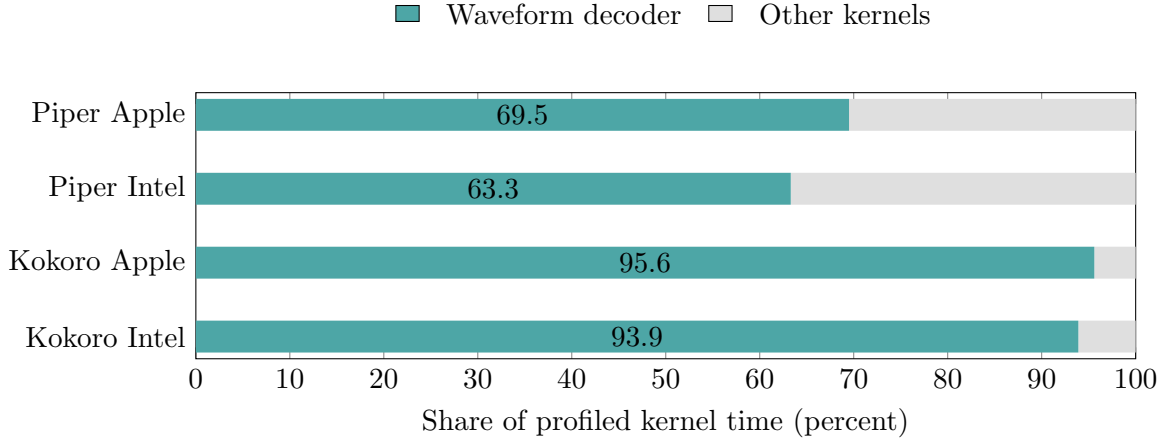
\begin{figure}[tbp]
\centering
\begin{tikzpicture}\begin{axis}[width=.88\textwidth,height=5.1cm,xbar stacked,bar width=12pt,
 xmin=0,xmax=100,ytick={0,1,2,3},yticklabels={Kokoro Intel,Kokoro Apple,Piper Intel,Piper Apple},
 xlabel={Share of profiled kernel time (percent)},legend style={at={(.5,1.2)},anchor=south,legend columns=2,column sep=8pt,draw=none},
 ytick style={draw=none},nodes near coords,point meta=x]
\addplot[fill=teal!70,draw=none] coordinates {(93.9,0) (95.6,1) (63.3,2) (69.5,3)};\addlegendentry{Waveform decoder}
\addplot[fill=gray!25,draw=none,nodes near coords={}] coordinates {(6.1,0) (4.4,1) (36.7,2) (30.5,3)};\addlegendentry{Other kernels}
\end{axis}\end{tikzpicture}

\caption{The decoder dominates the measured kernel time in both model families. The remainder includes all other kernels, not only the named stages in Table~\ref{tab:rq2}. These shares identify a useful target but do not measure the fraction of computation that can safely be skipped.}
\label{fig:cost}
\end{figure}

Per-node ONNX Runtime profiling identifies the waveform decoder as the largest measured kernel-time component in both systems (Table~\ref{tab:rq2}; Figure~\ref{fig:cost}). It is therefore a useful optimization target; profiling alone does not establish that it is gateable. Reported corpus RTF summaries are 0.018/0.023 for Piper and 0.156/0.194 for Kokoro on Apple/Intel respectively. The synthesis harness takes the median of three timed runs per waveform; the Intel result file reports mean RTF summaries. Kernel-time shares exclude unprofiled pipeline overhead and should not be substituted for end-to-end time shares without checking that overhead.

\begin{table}[htbp]
\centering
\caption{Share of ONNX kernel time by stage, from six utterances spanning the three domains. Listed stages do not exhaust all kernel time.}
\label{tab:rq2}
\small
\begin{tabular}{@{}llcc@{}}
\toprule
\textbf{Model} & \textbf{Stage} & \textbf{Apple M5 Pro} & \textbf{Intel Xeon 8488C} \\
\midrule
\multirow{4}{*}{Piper (VITS)} & HiFi-GAN decoder & \textbf{69.5\%} & \textbf{63.3\%} \\
 & normalizing flow & 10.9\% & 12.5\% \\
 & text encoder & 9.9\% & 12.8\% \\
 & duration predictor & 7.0\% & 8.1\% \\
\midrule
\multirow{2}{*}{Kokoro (StyleTTS2)} & iSTFTNet decoder & \textbf{95.6\%} & \textbf{93.9\%} \\
 & encoder + prosody & 4.3\% & 5.9\% \\
\bottomrule
\end{tabular}
\end{table}

\subsection{Deletion-policy comparison}
\label{sec:h3}
Table~\ref{tab:h3} reports mean PESQ for the Model-1 leave-self-out criterion and the spreading-matrix criterion with the $-16$\,dB offset. Comparisons between policies within each block use the same utterances and target budgets. The blocks use different subsets and band mappings, so differences between blocks cannot be attributed to the PAM alone. Energy has higher mean PESQ at all tabulated settings. The masking implementations also exceed random deletion at those settings.

For the spreading-matrix comparison, energy minus masking ranges from approximately 0.009 PESQ points (Piper, 10\% deletion) to 1.316 (Kokoro, 50\%). The logged paired Wilcoxon tests give $p=1.2\times10^{-6}$ for Piper at 10\% and approximately $1.2\times10^{-7}$ for the other five comparisons ($n=24$ paired utterances each). The six Model-1 comparisons have logged $p\approx5.7\times10^{-14}$ ($n=45$ each). These are exploratory, unadjusted tests on the fixed corpus, not evidence of perceptual equivalence or generalization to arbitrary voices. Mean differences should be read alongside their practical magnitude, particularly near the PESQ ceiling.

\begin{table}[htbp]
\centering
\caption{Mean wideband PESQ after deletion at matched target bin fractions. Model-1 uses 45 utterances per model; the spreading-matrix implementation uses 24 per model and a $-16$\,dB offset. Masking rankings use leave-self-out thresholds. Significance varies by comparison; see text.}
\label{tab:h3}
\small
\begin{tabular}{@{}llcccc@{}}
\toprule
\textbf{PAM} & \textbf{Model} & \textbf{Budget} & \textbf{energy} & \textbf{masking} & \textbf{random} \\
\midrule
\multirow{6}{*}{Model-1 (90 utt.)} & \multirow{3}{*}{Kokoro} & 0.1 & \textbf{4.57} & 4.32 & 2.30 \\
 & & 0.3 & \textbf{4.12} & 3.21 & 1.29 \\
 & & 0.5 & \textbf{2.98} & 1.71 & 1.09 \\
 & \multirow{3}{*}{Piper} & 0.1 & \textbf{4.64} & 4.56 & 2.32 \\
 & & 0.3 & \textbf{4.59} & 4.04 & 1.28 \\
 & & 0.5 & \textbf{4.40} & 2.69 & 1.09 \\
\midrule
\multirow{6}{*}{Spreading matrix (48 utt.)} & \multirow{3}{*}{Kokoro} & 0.1 & \textbf{4.59} & 4.39 & 2.87 \\
 & & 0.3 & \textbf{4.30} & 3.19 & 1.63 \\
 & & 0.5 & \textbf{3.42} & 2.10 & 1.22 \\
 & \multirow{3}{*}{Piper} & 0.1 & \textbf{4.64} & 4.63 & 2.76 \\
 & & 0.3 & \textbf{4.62} & 4.42 & 1.62 \\
 & & 0.5 & \textbf{4.46} & 3.60 & 1.19 \\
\bottomrule
\end{tabular}
\end{table}

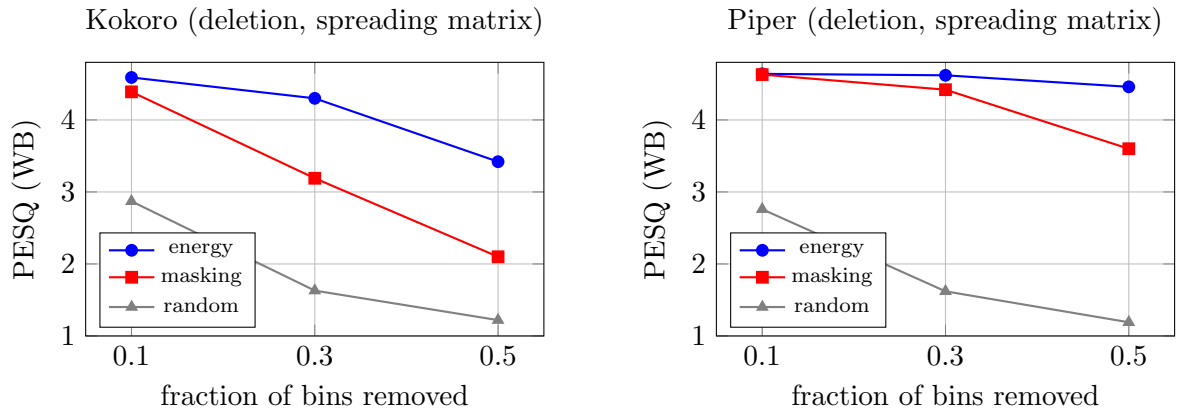
\begin{figure}[htbp]
\centering
\begin{subfigure}[t]{0.48\textwidth}
\centering
\begin{tikzpicture}
\begin{axis}[
    title={Kokoro (deletion, spreading matrix)},
    xlabel={fraction of bins removed},
    ylabel={PESQ (WB)},
    xmin=0.05, xmax=0.55, ymin=1, ymax=4.8,
    xtick={0.1,0.3,0.5},
    legend pos=south west, legend style={font=\scriptsize},
    grid=major, width=\textwidth, height=5.2cm,
]
\addplot[color=blue, mark=*, thick] coordinates {(0.1,4.59) (0.3,4.30) (0.5,3.42)};
\addlegendentry{energy}
\addplot[color=red, mark=square*, thick] coordinates {(0.1,4.39) (0.3,3.19) (0.5,2.10)};
\addlegendentry{masking}
\addplot[color=gray, mark=triangle*, thick] coordinates {(0.1,2.87) (0.3,1.63) (0.5,1.22)};
\addlegendentry{random}
\end{axis}
\end{tikzpicture}
\end{subfigure}
\hfill
\begin{subfigure}[t]{0.48\textwidth}
\centering
\begin{tikzpicture}
\begin{axis}[
    title={Piper (deletion, spreading matrix)},
    xlabel={fraction of bins removed},
    ylabel={PESQ (WB)},
    xmin=0.05, xmax=0.55, ymin=1, ymax=4.8,
    xtick={0.1,0.3,0.5},
    legend pos=south west, legend style={font=\scriptsize},
    grid=major, width=\textwidth, height=5.2cm,
]
\addplot[color=blue, mark=*, thick] coordinates {(0.1,4.64) (0.3,4.62) (0.5,4.46)};
\addlegendentry{energy}
\addplot[color=red, mark=square*, thick] coordinates {(0.1,4.63) (0.3,4.42) (0.5,3.60)};
\addlegendentry{masking}
\addplot[color=gray, mark=triangle*, thick] coordinates {(0.1,2.76) (0.3,1.62) (0.5,1.19)};
\addlegendentry{random}
\end{axis}
\end{tikzpicture}
\end{subfigure}
\caption{Deletion-policy comparison on 24 utterances per model. Energy achieves higher mean PESQ than the tested masking criterion, and both exceed random deletion. These are metric scores, not listener transparency judgments.}
\label{fig:h3}
\end{figure}

The result supports energy ranking as a strong baseline for this oracle task. It does not establish energy as a sufficient statistic of audibility, nor rule out improved masking models, different distortions, or perceptual verifiers applied to actual decoder errors. Fixed low-pass, frequency-only, and silence-removal baselines remain necessary to isolate the value of content adaptation.

\subsection{Noise-shaping comparison}
\begin{figure}[tbp]
\centering
\begin{minipage}{.48\textwidth}\centering\begin{tikzpicture}\begin{axis}[width=\textwidth,height=5.4cm,title={Kokoro},xmin=-46,xmax=-19,ymin=1.8,ymax=4.8,xlabel={Allocated NSR (dB)},ylabel={PESQ (WB)},xtick={-45,-35,-25,-20},legend pos=south west,legend style={font=\scriptsize},grid=major]
\addplot[color=teal!80!black,mark=*,thick] coordinates {(-45,4.6347) (-40,4.6204) (-35,4.5912) (-30,4.5401) (-25,4.4511) (-20,4.2853)};
\addlegendentry{Energy}
\addplot[color=red!65!black,mark=*,thick] coordinates {(-45,4.6315) (-40,4.6113) (-35,4.5679) (-30,4.4865) (-25,4.3169) (-20,4.049)};
\addlegendentry{Masking}
\addplot[color=gray,mark=*,thick] coordinates {(-45,4.181) (-40,3.8739) (-35,3.5087) (-30,3.0909) (-25,2.6515) (-20,2.2633)};
\addlegendentry{Uniform}
\end{axis}\end{tikzpicture}\end{minipage}\hfill
\begin{minipage}{.48\textwidth}\centering\begin{tikzpicture}\begin{axis}[width=\textwidth,height=5.4cm,title={Piper},xmin=-46,xmax=-19,ymin=1.8,ymax=4.8,xlabel={Allocated NSR (dB)},ylabel={PESQ (WB)},xtick={-45,-35,-25,-20},legend pos=south west,legend style={font=\scriptsize},grid=major]
\addplot[color=teal!80!black,mark=*,thick] coordinates {(-45,4.636) (-40,4.6244) (-35,4.5975) (-30,4.5351) (-25,4.4209) (-20,4.2352)};
\addlegendentry{Energy}
\addplot[color=red!65!black,mark=*,thick] coordinates {(-45,4.6046) (-40,4.5577) (-35,4.4643) (-30,4.2953) (-25,3.969) (-20,3.6364)};
\addlegendentry{Masking}
\addplot[color=gray,mark=*,thick] coordinates {(-45,3.9953) (-40,3.6259) (-35,3.1872) (-30,2.714) (-25,2.285) (-20,1.9579)};
\addlegendentry{Uniform}
\end{axis}\end{tikzpicture}\end{minipage}
\caption{Noise placement at equal allocated spectral noise power. Means are replotted from the released Model-1 noise sweep, with 15 utterances per model. Increasing NSR adds more noise. Energy-shaped and threshold-shaped noise both outperform uniform shaping in the reported means, while the energy-versus-masking gap is small at low noise levels. These are not deletion or compute budgets.}
\label{fig:noise}
\end{figure}
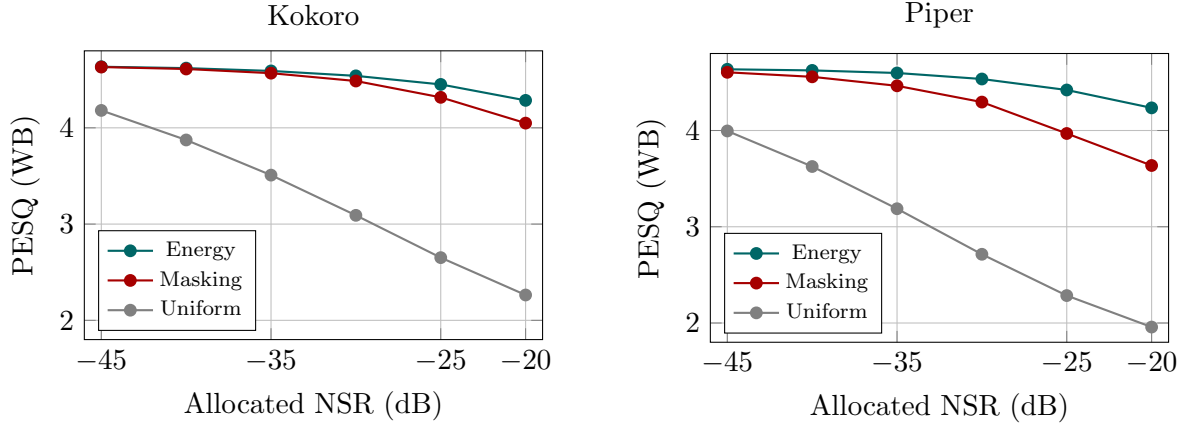

Figure~\ref{fig:noise} visualizes the noise-shaping comparison. At matched allocated noise power, energy shaping also has higher mean PESQ than Model-1 threshold shaping in the reported sweep from $-45$ to $-20$\,dB NSR (15 utterances per model). The size and statistical evidence of the difference vary. On Kokoro at $-45$\,dB, mean PESQ is 4.635 versus 4.632, with an unadjusted paired Wilcoxon $p\approx0.064$; this does not establish a difference at a 0.05 threshold. At $-20$\,dB, the corresponding means are 4.285 versus 4.049. For Piper, these pairs are 4.636 versus 4.605 at $-45$\,dB and 4.235 versus 3.636 at $-20$\,dB. Uniform noise shaping scores lower in each reported setting. The released noise sweep tests Model-1; it does not establish the same comparison for every PAM variant. STOI remains near its ceiling (approximately 0.993--1.000 across the logged conditions), so it adds little discrimination in this sweep.

\subsection{Masking calibration diagnostics}
\label{sec:probe}
The $\Delta$-probe injects noise shaped by the PAM threshold, scaled by $\Delta$\,dB, and records the resulting PESQ. This is a diagnostic of agreement with a chosen metric under a specified perturbation. A PESQ reference level, such as 4.4, is an engineering screen and is not a validated human-transparency threshold for these distortions. Threshold-shaped random noise also differs from a codec's actual reconstruction error; the reported LAME MP3 scores of 4.54--4.64 are contextual anchors, not proof that the PAM is perceptually calibrated.

\begin{figure}[tbp]
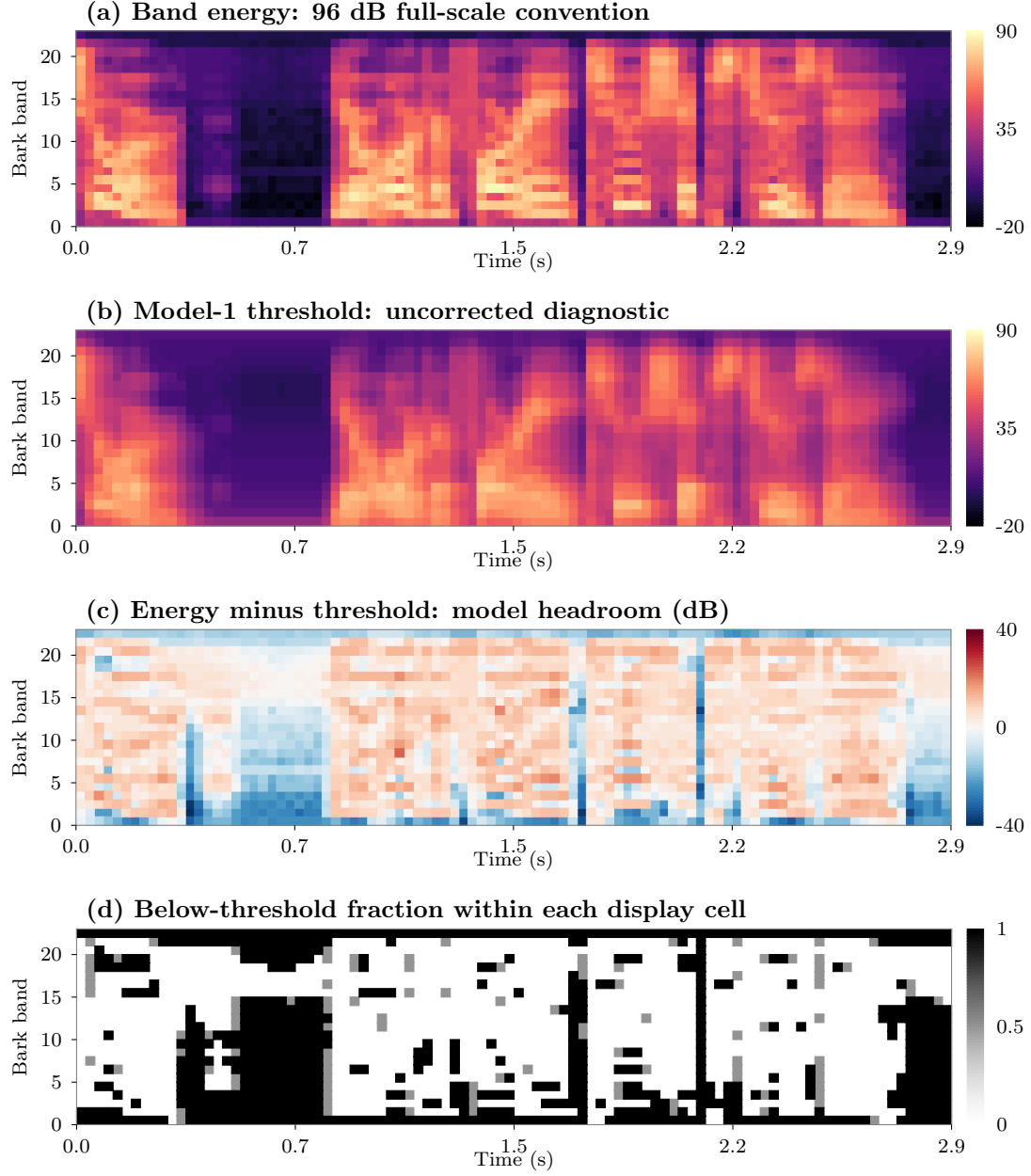

\centering

\caption{Masking diagnostic for the released, level-normalized Piper reference stimulus. Top to bottom: band energy, Model-1 threshold, model headroom, and the fraction of frames below threshold within each display cell. Positive headroom means above the model threshold; negative means below it. Dark cells in the last panel indicate model predictions, not proven inaudibility or wasted compute. The uncorrected Model-1 calibration is unreliable (Section~\ref{sec:probe}); the figure explains the analysis and must not be used to estimate savings.}
\label{fig:masking-map}
\end{figure}

The Model-1 calibration diagnostic indicates that substantially negative offsets, approximately $-30$ to $-40$\,dB, were needed to reach its metric screen. Band-summed noise maskers were identified as a source of excessive threshold estimates. Figure~\ref{fig:masking-map} shows four stages of this analysis: signal energy, threshold, headroom, and below-threshold decisions. It shows what the model computes, rather than where computation is known to be wasted. The corresponding high below-threshold corpus fractions remain diagnostic outputs in Appendix~\ref{app:masking}, not evidence of inaudibility or harvestable compute.

\begin{figure}[htbp]
\centering
\begin{tikzpicture}
\begin{axis}[
    xlabel={$\Delta$ (noise level relative to PAM threshold, dB)},
    ylabel={PESQ (WB)},
    xmin=-14, xmax=14, ymin=1, ymax=4.8,
    xtick={-12,-6,0,6,12},
    legend pos=north east, legend style={font=\scriptsize},
    grid=major, width=10cm, height=5.5cm,
]
\addplot[color=blue, mark=*, thick] coordinates {(-12,4.07) (-6,3.62) (0,2.83) (6,1.93) (12,1.48)};
\addlegendentry{spreading matrix, noise capped at signal}
\addplot[color=red, mark=square*, thick, dashed] coordinates {(-12,3.73) (-6,2.84) (0,1.90) (6,1.30) (12,1.15)};
\addlegendentry{spreading matrix, uncapped}
\addplot[color=black, thick, dotted, domain=-14:14] {4.4};
\addlegendentry{PESQ reference level (4.4)}
\end{axis}
\end{tikzpicture}
\caption{Reported spreading-matrix $\Delta$-probe on 18 utterances. Capped noise uses the signal magnitude as a cap before offset scaling. No displayed sample reaches the 4.4 reference level. The later $-16$\,dB setting lies outside this sweep; no unmeasured crossing is drawn.}
\label{fig:probe}
\end{figure}
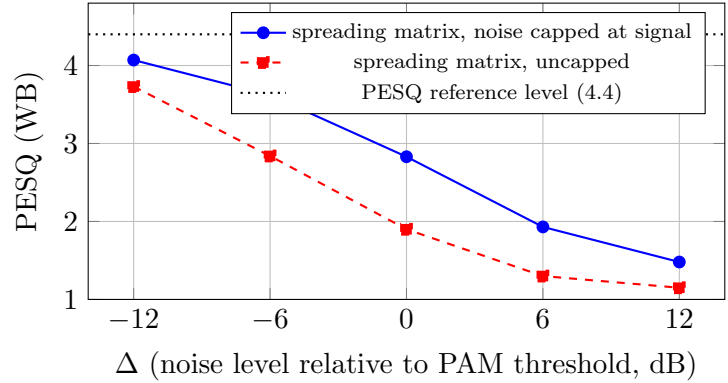

Figure~\ref{fig:probe} shows the spreading-matrix sweep. None of its sampled offsets reaches PESQ 4.4. The subsequent deletion harness fixes $\Delta=-16$\,dB, outside the plotted range; this figure does not establish a crossing at that value. With that offset, the reported mean below-threshold bin fractions are 0.291 for Piper and 0.116 for Kokoro (24 utterances per model). They are properties of the model and setting, not estimates of listener transparency. The corresponding logged fully-below-threshold frame fractions are zero on this subset.

The informal listening reports differ from the ordering and magnitude of some PESQ scores (Section~\ref{sec:abx}). Their sample size does not validate or invalidate PESQ for this distortion class. We use PESQ descriptively and leave both absolute perceptual calibration and listener agreement with policy rankings unresolved.

\subsection{Mel-resolution reconstruction diagnostic}
\label{sec:rq3}
The released mel-proxy script starts from the \emph{completed waveform's} power spectrum, averages it into 80 mel bands, redistributes that energy onto the linear-frequency grid, and recomputes Model-1 thresholds. Moreover, both sets of band decisions are compared against the original waveform's band energy. This is a diagnostic of sensitivity to spectral smoothing, not a predictor evaluated on pre-synthesis decoder features.

Across 30 waveforms, threshold RMSE is 2.78\,dB for Piper and 2.73\,dB for Kokoro, with decision agreement of 96.4\% and 95.9\%. Of cells classified as below threshold using the smoothed estimate, 11.4\% and 12.5\%, respectively, are above threshold under the unsmoothed calculation. These are disagreements between two computations, not measured rates of audible error. The reference Model-1 calculation itself has the calibration limitations discussed above. Actual predictive gating would require held-out evaluation using features available before waveform generation, along with the cost and quality effects of conservative decision margins.

\subsection{Dynamic-INT8 decoder experiment}
\label{sec:rq4}
\begin{figure}[tbp]
\centering
\begin{tikzpicture}\begin{axis}[width=.84\textwidth,height=5.2cm,ybar,bar width=20pt,
 symbolic x coords={Intel Xeon,Apple M5 Pro},enlarge x limits=.4,xtick=data,ymin=0,ymax=46,ylabel={Seconds for five utterances},
 legend style={at={(.5,1.03)},anchor=south,legend columns=2,column sep=8pt,draw=none},nodes near coords]
\addplot[fill=teal!70,draw=none] coordinates {(Intel Xeon,4.78) (Apple M5 Pro,3.87)};\addlegendentry{FP32}
\addplot[fill=orange!75,draw=none] coordinates {(Intel Xeon,38.56) (Apple M5 Pro,18.62)};\addlegendentry{Dynamic INT8}
\end{axis}\end{tikzpicture}

\caption{Fewer bits did not mean less time in this implementation. The tested dynamic-INT8 Kokoro decoder was slower on both CPUs. Values reproduce Table~\ref{tab:int8}; they do not compare every quantization scheme or runtime.}
\label{fig:quant}
\end{figure}
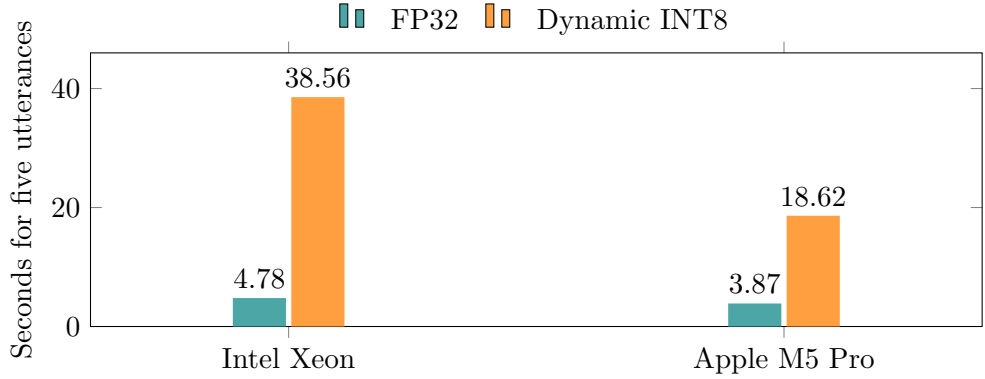

A separate experiment applies dynamic INT8 quantization to a Kokoro decoder split at five encoder/decoder boundary tensors. The unquantized split was reported as bit-exact against the monolithic model; supporting-data availability is described in Appendix~\ref{app:provenance}. Table~\ref{tab:int8} and Figure~\ref{fig:quant} show the measured timing comparison: this INT8 configuration is 8.1 times slower than FP32 on Intel and 4.8 times slower on Apple. It is a whole-decoder reduced-precision experiment, not an audibility-conditioned allocation policy.

Unsupported integer operators and conversion overhead are plausible contributors, but the timing table alone does not identify the cause. Establishing it requires inspecting the transformed graph and profiling individual kernels and conversions. Dynamic activation-parameter calculation can itself add inference overhead~\cite{ortquantization}. These results do not rule out static quantization, mixed precision, other runtimes, or architecture-specific kernels. The reported whole-model quantization attempt also shifted output durations; the tested decoder-only split preserves the upstream duration path. This observation does not make all whole-model draft approaches impossible.

\begin{table}[htbp]
\centering
\caption{Reported Kokoro decoder wall-clock for five utterances, best of three runs after warm-up, single process. Speedup is FP32 time divided by INT8 time; values below one indicate a slowdown.}
\label{tab:int8}
\small
\begin{tabular}{@{}lccc@{}}
\toprule
\textbf{Hardware} & \textbf{fp32} & \textbf{dynamic INT8} & \textbf{INT8 speed-up} \\
\midrule
Intel Xeon 8488C (AVX-VNNI) & 4.78\,s & 38.56\,s & $0.12\times$ \\
Apple M5 Pro & 3.87\,s & 18.62\,s & $0.21\times$ \\
\bottomrule
\end{tabular}
\end{table}

\subsection{Human evaluation}
\label{sec:abx}
\begin{figure}[tbp]
\centering
\begin{tikzpicture}\begin{axis}[width=.86\textwidth,height=6cm,xmin=0,xmax=1,ymin=-.6,ymax=5.6,
 xlabel={ABX accuracy},ytick={0,1,2,3,4,5},yticklabels={Random 30\%,Energy 90\%,Energy 75\%,Energy 60\%,Energy 50\%,Energy 30\%},
 grid=major,xtick={0,.25,.5,.75,1},legend style={at={(.5,1.02)},anchor=south,legend columns=2,column sep=8pt,draw=none}]
\addplot[gray,dashed,no marks] coordinates {(.5,-.6) (.5,5.6)};\addlegendentry{Chance (0.5)}
\addplot[orange!80!black,dotted,no marks] coordinates {(.6,-.6) (.6,5.6)};\addlegendentry{Proposed margin (0.6)}
\addplot[color=teal!80!black,no marks,thick] coordinates {(0.44,5) (0.97,5)};
\addplot[color=teal!80!black,only marks,mark=*,mark size=2.5pt] coordinates {(0.83333,5)};
\addplot[color=teal!80!black,no marks,thick] coordinates {(0.03,4) (0.56,4)};
\addplot[color=teal!80!black,only marks,mark=*,mark size=2.5pt] coordinates {(0.16667,4)};
\addplot[color=teal!80!black,no marks,thick] coordinates {(0.1,3) (0.7,3)};
\addplot[color=teal!80!black,only marks,mark=*,mark size=2.5pt] coordinates {(0.33333,3)};
\addplot[color=teal!80!black,no marks,thick] coordinates {(0.44,2) (0.97,2)};
\addplot[color=teal!80!black,only marks,mark=*,mark size=2.5pt] coordinates {(0.83333,2)};
\addplot[color=teal!80!black,no marks,thick] coordinates {(0.3,1) (0.9,1)};
\addplot[color=teal!80!black,only marks,mark=*,mark size=2.5pt] coordinates {(0.66667,1)};
\addplot[color=gray,no marks,thick] coordinates {(0.61,0) (1,0)};
\addplot[color=gray,only marks,mark=*,mark size=2.5pt] coordinates {(1,0)};
\end{axis}\end{tikzpicture}
\caption{The listening pilot is too small to establish a transparent deletion budget. Points show six-trial accuracies for one listener; lines show the reported Wilson 95\% intervals. The nonmonotonic pattern and wide intervals are visible directly. The 0.60 line is the archived proposed discrimination margin, not an established perceptual boundary.}
\label{fig:abx}
\end{figure}
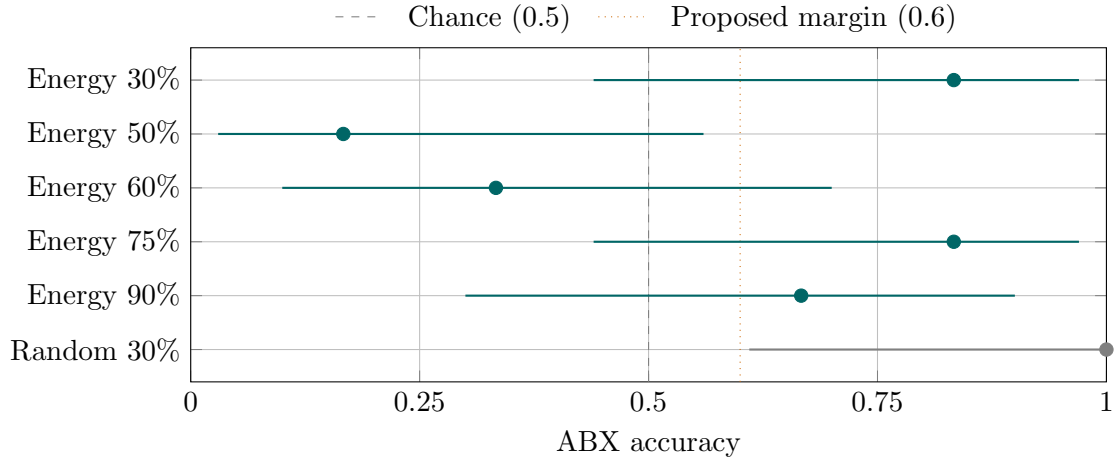

The author, who had heard related clips in an earlier informal session, completed a blind ABX pilot~\cite{clark1982abx} using eight utterances from the two models. There were two practice rounds, 30 energy-condition trials (six each at 30/50/60/75/90\% deletion), and six random-deletion anchor trials at 30\%. The listener identified all six anchors correctly, showing that this listener and playback setup detected that control degradation. Table~\ref{tab:abx} reports the pilot counts as accuracies and Figure~\ref{fig:abx} shows their uncertainty using Wilson intervals~\cite{wilson1927probable}.

\begin{table}[htbp]
\centering
\caption{Single-listener ABX pilot, six trials per condition. Chance accuracy is 0.50. Intervals describe this small pilot under a binomial model and do not support population-level transparency claims.}
\label{tab:abx}
\small
\begin{tabular}{@{}lccc@{}}
\toprule
\textbf{Condition} & \textbf{$n$} & \textbf{accuracy} & \textbf{Wilson 95\% CI} \\
\midrule
energy 30\% & 6 & 0.83 & [0.44, 0.97] \\
energy 50\% & 6 & 0.17 & [0.03, 0.56] \\
energy 60\% & 6 & 0.33 & [0.10, 0.70] \\
energy 75\% & 6 & 0.83 & [0.44, 0.97] \\
energy 90\% & 6 & 0.67 & [0.30, 0.90] \\
random 30\% (anchor) & 6 & 1.00 & [0.61, 1.00] \\
\bottomrule
\end{tabular}
\end{table}

Accuracy pooled across energy conditions is 17/30 (0.57), but this pooling does not establish a transparent budget. The condition-wise pattern is nonmonotonic: 5/6 correct at both 30\% and 75\%, compared with 1/6 at 50\%. Every energy-condition interval includes chance. Four extend above 0.60; the 50\% interval has an upper bound of 0.56. The latter numerical bound must not be presented as a confirmatory result: this was an exploratory, single, non-naive listener with six trials per condition. No removal budget is established as transparent.

An earlier informal session by the same listener reported ``same'', ``slight'', or ``clear'' differences, with a hidden reference and random-deletion control. Those observations are summarized in Appendix~\ref{app:pilot}; they neither independently corroborate the ABX pilot nor validate PESQ rankings. A larger study with independent listeners and prespecified analysis is required. The repository's \texttt{study/} directory provides the instrument and audio; Appendix~\ref{app:abx} distinguishes the existing pilot protocol from proposed future analysis.

\section{Implications for Adaptive Decoding}
\label{sec:roadmap}
Possible mechanisms include frame-level cheap/full decoder switching, band-limited synthesis, and students trained to condition capacity on predicted output sensitivity. These are research directions, not implemented results. They must preserve state and boundary continuity, make decisions from available inputs, and include routing, verification, and fallback costs in end-to-end timing.

For an illustrative cost model, let $d$ be the decoder's fraction of baseline time and suppose its cost falls in direct proportion to an output fraction $f$ skipped, with no additional overhead. Overall speedup is then
\[
 S=\frac{1}{(1-d)+d(1-f)}=\frac{1}{1-df}.
\]
Substituting the Intel kernel-time shares (approximately 0.63 and 0.94) and $f=0.5$ gives roughly 1.5 and 1.9 times speedup. These are conditional kernel-time projections, not end-to-end measurements or general upper bounds for future architectures. The assumptions that bin removal corresponds to skipped decoder cost and that kernel shares represent the full pipeline have not been demonstrated~\cite{amdahl1967validity}.

\section{Related Work}
Perceptual codecs allocate bits using auditory models~\cite{iso11172,painter2000perceptual,bosi2003introduction}. Neural audio work also uses psychoacoustic loss weighting and multiband reconstruction~\cite{zhen2018psychoacoustically,zhen2020calibration,muffin2025}. Cui et al.~\cite{cui2020subband} explicitly use auditory sensitivity to motivate greater modeling resources for a low-frequency synthesis subband. Thus perceptually motivated nonuniform synthesis resources are an existing idea. Our experiments examine content-dependent post-synthesis perturbation placement as a possible prerequisite for inference allocation; they do not implement dynamic inference allocation.

Adaptive computation includes early exits and learned routing~\cite{teerapittayanon2016branchynet,schuster2022calm,han2022dynamic,raposo2024mixture}, with speech-representation examples~\cite{daisy2024}. Synthesis-side efficiency includes speculative decoding~\cite{leviathan2023fast,chen2023accelerating,ssd2025}, dynamic chunk prediction~\cite{dcar2025}, and scheduled early exit~\cite{spark2026}. Other approaches improve synthesis architecture or reduce sampling steps~\cite{valin2019lpcnet,ren2021fastspeech2,zheng2025epss}. These methods motivate comparisons for a future implemented allocator; this paper makes no exhaustive priority claim over adaptive speech synthesis.

In vision, foveated rendering and reconstruction allocate work according to visual sensitivity~\cite{guenter2012foveated,patney2016foveated,kaplanyan2019deepfovea}, with related work on spatially adaptive diffusion~\cite{foveateddiffusion2026}. Perceptually guided acoustic rendering is another relevant precedent~\cite{peng2025foveation}. The proposed TTS direction draws on these ideas, while its computational benefit remains untested here.

\section{Limitations and Next Experiments}
The corpus is small, fixed, and limited to two model configurations, with Hinglish processed through English voices. The listening evidence comes from one non-naive author. No population-level transparency, independent perceptual policy ranking, conditional-compute implementation, or end-to-end speedup is established. The masking calculations depend on playback-level conventions and implementation choices; leave-self-out thresholds do not resolve all simultaneous-removal effects.

PESQ and STOI score resampled 16\,kHz audio, whereas native-rate bin budgets extend above 8\,kHz. The next controls should restrict degradation and scoring to a common bandwidth and compare energy ranking with low-pass, frequency-only, and silence-removal policies. They should report removed signal energy and actual removed-bin fractions alongside target budgets. Separate calibration and evaluation sets are needed for any selected threshold setting.

The paired tests describe this corpus; they are unadjusted for multiple comparisons, and the displayed mean-score curves do not include uncertainty intervals. More voices, texts, random placements, and paired uncertainty estimates are needed to assess stability. Timing uses only three repeats and, for decoder microbenchmarks, selects the best repeat; it does not characterize latency tails or production throughput. The source release documents substantial parts of the workflow but does not establish that every manuscript-only measurement is independently reproducible from the committed files.

A future multi-listener study should prespecify its population, playback conditions, screening, equivalence margin, sample size, and treatment of repeated listeners and utterances. A subsequent compute experiment should compare an implemented allocator against a full decoder and relevant simpler baselines under the same quality evaluation, with all overhead included.

\section{Conclusion}
Hearing's blind spots suggest a useful question for synthesis: which details deserve computation? In two neural TTS systems, energy-ranked spectral deletion yields higher mean PESQ than the tested masking criteria and random deletion at matched target bin budgets. This supports energy ranking as a strong baseline for oracle perturbation studies. Calibration diagnostics, an inconclusive author-only listening pilot, and a slow dynamic-INT8 decoder configuration identify limits of the current evidence. Perceptual transparency and conversion of output sensitivity into inference savings remain open. The released experiments provide a basis for testing those questions without treating removed spectral output as measured compute reduction.

\section*{Acknowledgments}
The reported experiments used an Apple M5 Pro and an AWS \texttt{c7i.2xlarge}. We thank an external reviewer for feedback on claim scope.

\emph{Use of AI tools.} Anthropic Claude assisted with portions of the experiment-harness code, analysis scripts, manuscript drafting, and identification of candidate related work. OpenAI Codex assisted with manuscript preparation, including checks against released scripts and logs, methodological descriptions, reference-metadata checks, and visualizations of released audio and reported results. These tools are not authors and their involvement does not constitute independent experimental validation. The author is responsible for the final content, citations, and interpretation.

\appendix
\section{Diagnostic Outputs from the Original Masking Analysis}
\label{app:masking}
The first four rows of Table~\ref{tab:h1} show Model-1 outputs at the 96\,dB full-scale convention (45 utterances per model). Because subsequent diagnostics found excessive thresholds, these values are not estimates of inaudible fractions or feasible frame skipping. The last row comes from a different implementation and subset with a chosen offset, so the table is not a controlled before/after calibration comparison. Figure~\ref{fig:masking-map} shows the four-panel diagnostic computed from the released level-normalized listening stimulus and labeled as a model calculation rather than a validated audibility map.

\begin{table}[htbp]
\centering
\caption{Diagnostic model outputs only. Original Model-1 rows: 45 utterances per model. Spreading-matrix row: 24 per model. No row establishes human transparency or saved compute.}
\label{tab:h1}
\small
\begin{tabular}{@{}lcc@{}}
\toprule
\textbf{Metric} & \textbf{Piper} & \textbf{Kokoro} \\
\midrule
Model-1: bins below threshold & 0.977 & 0.977 \\
Frames with $\geq$95\% of bins below threshold & 0.90--0.91 & 0.86--0.88 \\
Frames fully below threshold & 0.10--0.16 & 0.14--0.15 \\
Signal energy below threshold & $\approx$0.16 & $\approx$0.18 \\
Spreading matrix: bins below threshold ($\Delta=-16$\,dB) & 0.291 & 0.116 \\
\bottomrule
\end{tabular}
\end{table}

\section{Earlier Informal Listening Observations}
\label{app:pilot}
Table~\ref{tab:pilot} summarizes the earlier subjective reports from the same author, listening on speakers in a quiet setting. ``Same'' means the listener reported no difference in that informal comparison. It is not an equivalence-test outcome, and these observations do not justify a general 60--75\% transparency claim.

\begin{table}[htbp]
\centering
\caption{Original informal single-listener observations (speakers, quiet). Entries are reports of same/slight/clear difference, not validated transparency labels.}
\label{tab:pilot}
\small\setlength{\tabcolsep}{3.5pt}
\begin{tabular}{@{}lcccc@{}}
\toprule
\textbf{Condition} & \textbf{Piper} & \textbf{Kokoro} & \textbf{PESQ (Piper)} & \textbf{PESQ (Kokoro)} \\
\midrule
energy 60\% / masking 60\% & same / same & same / same & 4.48 / 3.06 & 3.08 / 1.80 \\
energy 75\% / masking 75\% & same / same & same / slight & 3.70 / 2.53 & 2.01 / 1.40 \\
energy 90\% / masking 90\% & same / slight & slight / clear & 1.88 / 1.64 & 1.24 / 1.14 \\
random 30\% (anchor) & clear & clear & 1.58 & 1.74 \\
hidden reference & same & same & n/a & n/a \\
\bottomrule
\end{tabular}
\end{table}

\section{Reproducibility and Provenance}
\label{app:provenance}
The recorded software environment is Python 3.12.13; ONNX Runtime 1.28.0 (Apple) and 1.29.0 (Intel); piper-tts 1.6.1; kokoro-onnx 0.5.0; pesq 0.0.4; pystoi 0.4.1; and onnx 1.22.0. Model files are \texttt{en\_US-lessac-medium.onnx}, \texttt{kokoro-v1.0.onnx}, and \texttt{voices-v1.0.bin}. The reported random seed is 1234. These versions describe the experimental runs.

The scripts and archived results correspond to repository commit \href{\releaseurl}{\texttt{b0394458}} (the hyperlink identifies the full immutable revision). The corpus is in \texttt{scripts/corpus.py}; deletion comparisons are in \texttt{results/h3\_proxy3.json} and \texttt{results/h3\_pam2.json}, with corresponding run logs; noise comparisons are in \texttt{results/h3\_noise.json}; and the mel diagnostic is in \texttt{scripts/rq3\_mel\_proxy.py} and its result file. The Intel timing comparison is in \texttt{results/x86\_results.json}. The amendment is in \texttt{reports/amendment-1.md}. A commit-pinned document makes its contents identifiable; it does not establish that the document preceded the experiments it describes.

The ABX counts, earlier informal listening observations, Apple decoder microbenchmark, and spreading-matrix probe are aggregate results recorded in the manuscript. Their complete trial-level records or generating scripts were not verified in the pinned release. The unquantized decoder split's reported bit-exactness was not independently rerun. These gaps limit independent reproduction of those results. During manuscript preparation, descriptions and aggregate values were checked against available scripts and logs; this process did not include new synthesis, timing, or listening trials.

The spectrograms use the released \texttt{piper\_agent0\_ref.wav}, \texttt{piper\_agent0\_e30.wav}, and \texttt{piper\_agent0\_anchor.wav} listening stimuli. Display spectra use a 1024-sample Hann window and 512-sample hop, average power into 96 time columns and 64 frequency rows, and share a decibel reference equal to the original stimulus's peak STFT power. Figure~\ref{fig:masking-map} recomputes \texttt{pam.analyze} on the level-normalized reference with the original 96\,dB full-scale convention and forward masking enabled. Its first three panels average decibel values in display-time bins; its final panel shows the fraction of constituent frames classified below threshold. The conceptual masking diagram is illustrative; the other charts visualize existing reported results. These visualizations do not constitute new synthesis, timing, or listening experiments.

\section{Listening Protocol and Proposed Follow-Up}
\label{app:abx}
The pilot instrument uses level-normalized 16-bit PCM, two practice rounds with feedback, 30 energy-condition rounds (six per budget), and six anchor rounds, shuffled. An utterance--condition pair is not repeated within a participant. Assignments to A/B and X are randomized per round. The archived protocol specifies exclusion for anchor accuracy below 5/6 or median test-round response time below 1.5\,s, and states consent, anonymity, and withdrawal provisions.

The archived analysis pools trials by condition, computes Wilson 95\% intervals and one-sided binomial tests against 0.5, and proposes a transparency rule requiring both $p\geq0.05$ and an upper interval bound at most 0.60. This describes the historical protocol; it is not endorsed here as a sufficient population-level analysis. Failure to reject chance is not evidence of equivalence, and pooling ignores potential dependence within listeners and utterances. The pilot was not a confirmatory application of this future-study rule.

Before new recruitment, a revised analysis plan should define a practical discrimination margin and test whether the upper confidence bound lies below it using an approach that accounts for listener and utterance dependence, such as a prespecified hierarchical model or suitable clustered resampling. Sample size should be justified for that design, rather than treating all trials as independent. Per-model results, control performance, exclusions, and all budget comparisons should be reported. These recommendations concern future data collection; they were not preregistered decisions governing the pilot.
\end{document}